\documentclass[4pt,a4paper,twoside,twocolumn,american,prd,nofootinbib,superscriptaddress]{revtex4-1}
\usepackage{graphics}
\usepackage[pdftex]{graphicx}
\usepackage{amsmath, amssymb}
\usepackage{color}
\usepackage[usenames,dvipsnames]{xcolor}
\usepackage{setspace}
\usepackage{latexsym}
\usepackage{amssymb}
\usepackage{graphicx,color}
\usepackage{epstopdf}
\usepackage{makeidx} 
\usepackage{textcomp}
\usepackage{lmodern}
\usepackage[T1]{fontenc}
\usepackage[utf8]{inputenc}
\usepackage{color}
\usepackage{babel}
\usepackage{booktabs}
\usepackage{units}
\usepackage{amsmath}
\usepackage{amssymb}
\usepackage{esint}
\usepackage{subfigure}
\pdfpageheight\paperheight
\pdfpagewidth\paperwidth

\usepackage[unicode=true,pdfusetitle,
 bookmarks=true,bookmarksnumbered=false,bookmarksopen=false,
 breaklinks=false,pdfborder={0 0 0},backref=false,colorlinks=true]
 {hyperref}
\hypersetup{
 citecolor=blue,filecolor=blue,linkcolor=blue,urlcolor=blue}
\usepackage{aas_macros}

\def\be{\begin{equation}}
\def\ee{\end{equation}}
\def\ba{\arraycolsep .1em \begin{eqnarray}}
\def\ea{\end{eqnarray}}

\def\r{r_{T}}
\def\mpl{{M_{\rm pl} }}

\def\gef{g^{E}_{\mu\nu}}
\def\gej{g_{\mu\nu}}

\def\eq#1{(\ref{#1})}

\def\Eq#1{Eq.~(\ref{#1})}
\def\fig#1{fig.~(\ref{#1})}

\def\s0#1#2{\mbox{\small{$ \frac{#1}{#2} $}}}
\def\0#1#2{\frac{#1}{#2}}

\newcommand{\bit}{\begin{itemize}}
\newcommand{\eit}{\end{itemize}}
\newcommand{\planck}{{\em Planck} }
\begin{document}

\title{Asymptotically safe inflation from quadratic  gravity}

\author{Alfio Bonanno}
\affiliation{INAF, Osservatorio Astrofisico di Catania, via S. Sofia 78, I-95123 Catania, Italy}
\affiliation{INFN,  Sezione di Catania,  via S. Sofia 64, I-95123, Catania, Italy.}
\author{Alessia Platania}
\affiliation{INAF, Osservatorio Astrofisico di Catania, via S. Sofia 78, I-95123 Catania, Italy}
\affiliation{INFN,  Sezione di Catania,  via S. Sofia 64, I-95123, Catania, Italy.}
\affiliation{Universit\`a di Catania, via S. Sofia 63, I-95123 Catania, Italy}

\begin{abstract}
Asymptotically Safe  theories of gravity have recently received much attention. 
In this work we discuss a class of inflationary models derived from quantum-gravity 
modification of quadratic gravity according to the induced scaling around the non-Gaussian fixed point at very high energies.
It is argued  that the presence of a three dimensional ultraviolet critical surface generates operators of non-integer power 
of the type $R^{2-\theta/2}$ in the effective Lagrangian, where $\theta>0$ is a critical exponent.
The requirement of a successful inflationary model in agreement with  the recent \planck 2015 data 
puts important constraints on the strenght of this new type of couplings.  
\end{abstract}

\maketitle

\section{Introduction}
If the inflationary scenario is correct, then it is possible that quantum gravitational phenomena could be 
detected in anisotropy experiments of the cosmic microwave background radiation 
and in observations of the large scale structure of the Universe \cite{hh}. 

Although  a consistent quantization of the gravitational field is still lacking,   
it is assumed that quantizing the matter fields and the gravitational  field  on a classical
background should  be  sufficient  to explain the generation of the initial spectrum of perturbations. 

In recent times a promising theory of  quantum gravity has appeared in the framework  of the {\em asymptotic safety} (AS) proposal. 
Its central idea, as first discussed by Weinberg \cite{we79}, 
is to define the continuum limit  around a non-Gaussian fixed point (NGFP)
characterized by an ultraviolet (UV) critical manifold of finite dimensionality 
\cite{martin,dou,so99,la02,resa02b,resa02,daniel04,bore05,co06,co09,bema09,bema10,
mare11,ma11,ast13,beca,bogu,dimo13,begu14,begu,demme,dimo15}.

One of the most striking consequences of this approach is the fact that the fundamental theory
seems to be rather different from the classical gravity based on the Einstein-Hilbert action.
In the infinite cutoff limit, it is characterized by a vanishing  Newton's constant 
which is therefore {\em antiscreened} at high energies.
Classical gravity can only be recovered as a low-energy effective theory at some intermediate 
scale and {it does not} define a fundamental theory.  Although we do not know yet the structure 
of the fundamental Lagrangian, all the investigations performed so far, considering general $f(R)$
truncations or more complicated tensorial structure like $R_{\mu\nu}R^{\mu\nu}$, have  confirmed the 
finite dimensionality of the UV critical surface (see \cite{nie, resa13, nink} for reviews and also \cite{ohpe}
for a recent investigation within the  $f(R)$ truncation).

In the simple case of two relevant directions, as in the case of  {\em quantum} Einstein-Hilbert gravity,
several investigations have focused on the implications of the running of the Newton's constant 
in models of the Early Universe.
In \cite{cosmo1,irfp,resa05,cosmo2,conpe} 
it has been shown that the renormalization-group induced evolution of the Newton's constant and
cosmological constant can provide a consistent cosmic history of the Universe 
from the initial singularity to the superaccelerated expansion. 

In particular in \cite{cosmo2} it has been argued that the scaling properties of the 
2-points correlation function of the graviton near the NGFP
induce a scale invariant spectrum of the primordial perturbations, characterized by a 
spectral index $n_s$ which, to  a very good approximation, must satisfy $n_s \approx 1$. 

Subsequent investigations have tried to produce successful models of inflation by considering 
an extended  structure of the effective Lagrangian near the NGFP \cite{weinberg10}, 
but it turns out that the use of an ``optimal cutoff'' might result in a fine tuning problem  \cite{tye10,hong12}. 
In \cite{cai11} the running of the gravitational and cosmological constants has been described
in terms of a Jordan-Brans-Dicke model with a vanishing 
Brans-Dicke parameter, while 
the viability of the AS
scenario in models where the inflaton is the Higgs field has been discussed 
in \cite{cai13,xi14}.

An effective Lagrangian encoding  the leading quantum gravitational effect near the NGFP
has been proposed in \cite{alfio12} where  
a  RG-improvement  of the linearized $\beta$-functions has been 
performed using the idea that the relevant cutoff in this situation
is provided by the local curvature, i.e. $k^2\sim R$ 
(see also \cite{hi12}).
In a recent investigation a similar approach  has been 
successfully applied to study the quantum gravity modifications of the
Starobinsky model \cite{co15} (see also \cite{ri14,ri15,cosa15,inff1,inff2,inff3} for other examples
of quantum deformed quadratic gravity inflationary models).

In this work we extend the analysis of \cite{alfio12}
by including the additional relevant direction produced by  the $R^2$ operator
on  the UV critical surface.

It will be shown that the requirement of a successful inflationary model
in agreement with the recent \planck 2015 data release of CMB anisotropy
puts significant constraints on the renormalized flow generated 
by the $R^2$ term. In particular the dynamics of the inflaton after the inflationary phase
can be rather different from the well known $R+R^2$ Starobinsky model. 

The rest of the paper is organized as follows: 
In Section II we introduce the basic formalism and obtain the effective field theory description
in terms of a $f(R)$ theory. In Section III we compute the spectral index $n_s$ for the primordial 
perturbations and the tensor-to-scalar ratio $r$ using a conformal mapping to the Einstein frame. 
In Section IV the oscillatory phase after inflation is discussed, and Section V is devoted to the conclusions. 

\section{Basic formalism}
Let us consider  the quadratic gravity Lagrangian
\be
L_k=\frac{k^2}{16 \pi g_k}(R-2\lambda_k k^2)  -\beta_k R^2 
\label{r2}
\ee
where $k$ is a running energy scale and $g_k$, $\lambda_k$ and $\beta_k$ are dimensionless running 
coupling constants whose infinite momentum limit is controlled by a
NGFP, i.e. $\lim_{k\rightarrow\infty}(g_k, \lambda_k, \beta_k)
=(g_\ast, \lambda_\ast, \beta_\ast)\not = (0,0,0)$  \cite{la02,re12}:
does this running show up at the level of predictions for a specific inflationary model?

In the case at hand the qualitative behavior of the UV critical manifold of the {\em quantum} theory defined by \eq{r2} at the NGFP
is rather simple, as its continuum limit can be described only by three relevant couplings.
In particular  there exist trajectories that emanate from the NGFP and possess a long classical regime 
where the effective action is approximated by the standard Einstein-Hilbert action \cite{re12}.
On the contrary the quantitative details of the renormalized flow around the NGFP are still rather uncertain. 
In fact, not only the precise location of the NGFP is  regulator dependent (as expected), but also the value of the critical exponents
depends on the truncation strategy employed to solve the flow equation \cite{co09,fa14}. 
Moreover recent investigations based on unimodular gravity \cite{ast15}, and general arguments \cite{bel}
suggest that the critical exponents are indeed real \cite{nink15,peva15}. 

For these reasons we would like to find an approximation to the renormalized flow  which encodes the general 
qualitative feature of the scaling around the NGFP and assumes real critical exponents. 
We  thus approximate the running of $\lambda_k$ with its tree-level scaling 
$\lambda_k\sim c_0\,k^{-2}$ where $c_0$ is a dimensionful constant,  
and decouple the running of $g_k$  from the running of  $\beta_k$. 
This latter approximation is justified by the impressive stability of the critical exponents for the Newton's constant
against the inclusion of higher order truncations as shown in several investigations \cite{co09}, 
but it has also the important advantage to allow for an analitic expression of the flow 
in our model. 
In fact, under these assumptions the renormalized flow thus reads \cite{bore05}
\ba
\label{rug}
&&g_k = \frac{ 6 \pi c_1 k^2 }{6\pi\mu^2 + 23 c_1 (k^2-\mu^2)}\\[2mm]
&&\beta_k = \beta_{\ast}+b_0 \left (\frac{k^2}{\mu^2} \right )^{-\frac{\theta_3}{2}}
\label{rub}
\ea
where $\mu$ is an infrared renormalization point, $c_1=g_k(k=\mu)$ (see \cite{bore05} for details). 
According to \cite{la02} $\beta_\ast = \beta_k(k\rightarrow\infty)\simeq 0.002$, while $b_0$ is a free parameter obtained 
by the linearization of the RG flow around the NGFP, and $\theta_3$ is the critical exponents for the $R^2$ coupling.  

It is important to stress that, as long as $c_1< 6\pi/23$ 
the running described by \Eq{rug} smoothly interpolates between  the Gaussian fixed point (GFP) and the NGFP,  $g_\ast=g_k(k\rightarrow\infty)=6\pi/23$
and therefore it captures the qualititive features of the flow described in \cite{re12}. 
The constants $b_0$, $c_0$ and $c_1$ depend on the physical situation at hand and must be fixed 
by confronting the model with observations: in principle these are the only 
free parameters of our theory corresponding to the three relevant directions of the UV critical surface of the action \eq{r2}.
In other words, by changing $c_1$, $c_0$ and $b_0$ it is possible to explore various RG trajectories all ending at the 
NGFP.  The relevant question is if it possible to actually constraints these numbers, in particular
the value of $b_0$.

As argued  in \cite{alfio12,saltas12,co15}   
by substituing \eq{rug}, \eq{rub} and $\lambda_k=c_0/k^2$ in \eq{r2}, 
a renormalization group improved  effective Lagrangian
can be obtained  by the scale identification $k^2\rightarrow \xi R$, where  $\xi$ is positive number. 
It must be stressed that  the general structure of the resulting RG-improved effective Lagrangian agrees very well
with the high-curvature solution of the fixed point equation for a generic
asymptotically safe $f(R)$ theory, as it emerges from the analysis of \cite{dimo13} and \cite{alfio12}.

At last, we obtain  the following RG-improved action 
\be
\label{eff}
S=\frac{1}{2\kappa^2}\int d^4 x \sqrt{-g} \;\left [ R + \alpha R^{2-\frac{\theta_3}{2}}+ \frac{R^2}{6 m^2} -\Lambda \right ]
\ee
where $\mu$ is chosen so that
$\kappa^2= 8 \pi G_N = 48 \pi^2 c_1/(6\pi\mu^2-23(\mu^2+2\xi c_0)c_1)\equiv1/\mpl^2$. Moreover  
$\Lambda=\mu^2 c_0(6\pi -23 c_1)/(6\pi \mu^2 -23 (\mu^2+2\xi c_0) c_1)$ and 
$\mpl^2/m^2=12(23\xi/(96\pi^2-\beta_{\ast}))$; 
in particular if  $\xi> 96 \pi^2 \beta_{\ast}/23$ and $c_0 < \mu^2 (6\pi-23 c_1)/46\xi c_1$, then
 $1/m^2$ and $\Lambda$ are positive definite.

On the other hand 
\be
\alpha = - 2 \mu^{\theta_3}b_0 /\mpl^2
\ee
which only depends on $\theta_3$ and $b_0$, but not on the fixed point values.
Concerning $\theta_3$ the numerical
evidence accumulated so far  has shown that its value is rather stable against the introduction
of higher order truncation in the flow equation \cite{fa14}, as it should be 
expected for  a critical exponent. 
On the contrary $b_0$ is by construction a non-universal quantity whose value cannot 
be determined by the RG group. It  labels a
specific trajectory emanating from the fixed point and its actual value should be
determined by matching with  a low energy observable.

Let us now introduce an auxiliary field $\varphi$ defined via
\be
\label{infl}
\varphi(\chi) \equiv 1+\alpha \left(2-\frac{\theta_3}{2}\right)\chi^{1-\frac{\theta_3}{2}} + \frac{\chi}{3 m^2}
\ee
In principle this relation can always be inverted at least locally, provided $\varphi_{,\chi} \not = 0$
so that 
\eq{eff} is equivalent to 
\ba
\label{effu}
S=&&\frac{1}{2\kappa^2}\int d^4 x \sqrt{-g} \:\bigg[ \varphi R - \left(\varphi -1\right)\chi(\varphi)  \nonumber  \\[0.15cm]
&&- \alpha \bigg(2-\frac{\theta_3}{2}\bigg)\chi(\varphi)^{1-\frac{\theta_3}{2}} + \frac{\chi(\varphi)}{3 m^2}\bigg]
\ea
In practice, due to its non-linearity,  the task of inverting \Eq{infl} can be very difficult and one must often resort to numerical 
methods.  In our case, according to the analysis of \cite{co09}, as $\theta_3$
is rather close to unity we can safely set $\theta_3 = 1$ for any practical calculation (as it can be
numerically checked).
In this case we explicitly obtain the two branches
\ba
&&\chi_{\pm}= \frac{3}{8} \Big(27 \alpha ^2 m^4+8 m^2 \varphi -8 m^2 \nonumber\\[0.1cm]
&& \pm \, 3 \sqrt{3}\, \sqrt{27 \alpha^4 m^8+16 \alpha ^2 m^6 (\varphi -1)}\Big)
\ea
with the reality condition $\chi \ge 1-27 m^2 \alpha^2/16$. 

Using these solutions we can obtain a canonically coupled scalar field by introducing a conformally related metric $\gef$
by means of $\gef=\varphi \gej$ with $\varphi = e^{\sqrt{2/3}\kappa \phi}$.
In the Einstein frame\footnote{Note that, as discussed by \cite{calmet} the field redefinition employed 
in going from the Jordan to the Einstein frame involves new additional contributions in the path-integral
arising from the Jacobians. In investigations including the presence of matter field in the starting
Lagrangian, these new terms cannot be neglected.}\cite{defe,cade} the action thus reads
\be
\label{effe}	
S=\int d^4 x \sqrt{-g_E} \left [ \frac{1}{2\kappa^2} R_E -\frac{1}{2} g_E^{\mu\nu} \partial_\mu \phi \partial_\nu \phi -V_{\pm}(\phi) \right ] 
\ee
where
\ba
\label{effv}
&&V_{\pm}(\phi)=\frac{m^2 \mathrm{e}^{-2\sqrt{\frac{2}{3}}\kappa\phi}}{256\kappa^2}\bigg\{192\left(\mathrm{e}^{\sqrt{\frac{2}{3}}\kappa\phi}-1\right)^2-3\alpha^4+128\Lambda\nonumber\\[0.15cm]
&&-3\alpha^2\Big(\alpha^2+16\,\mathrm{e}^{\sqrt{\frac{2}{3}}\kappa\phi}-16\Big)\mp 6\alpha^3\sqrt{\alpha^2+16\,\mathrm{e}^{\sqrt{\frac{2}{3}}\kappa\phi}-16}\nonumber\\[0.15cm]
&&-\sqrt{32}\alpha\Big[\left(\alpha^2+8\mathrm{e}^{\sqrt{\frac{2}{3}}\kappa\phi}-8\right)\pm\alpha\sqrt{\alpha^2+16\mathrm{e}^{\sqrt{\frac{2}{3}}\kappa\phi}-16}\Big]^{\frac{3}{2}}\bigg\}\nonumber\\
\ea
and we have measured $\alpha$ and $\Lambda$ in units of the scalaron mass $m$ by means of the rescaling $\alpha \rightarrow \alpha / 3 \sqrt{3} m$
and $\Lambda \rightarrow \Lambda m^2$, so that both $\alpha$ and $\Lambda$ are dimensionless numbers.

\begin{figure}[t]
\centering
\includegraphics[width=8cm]{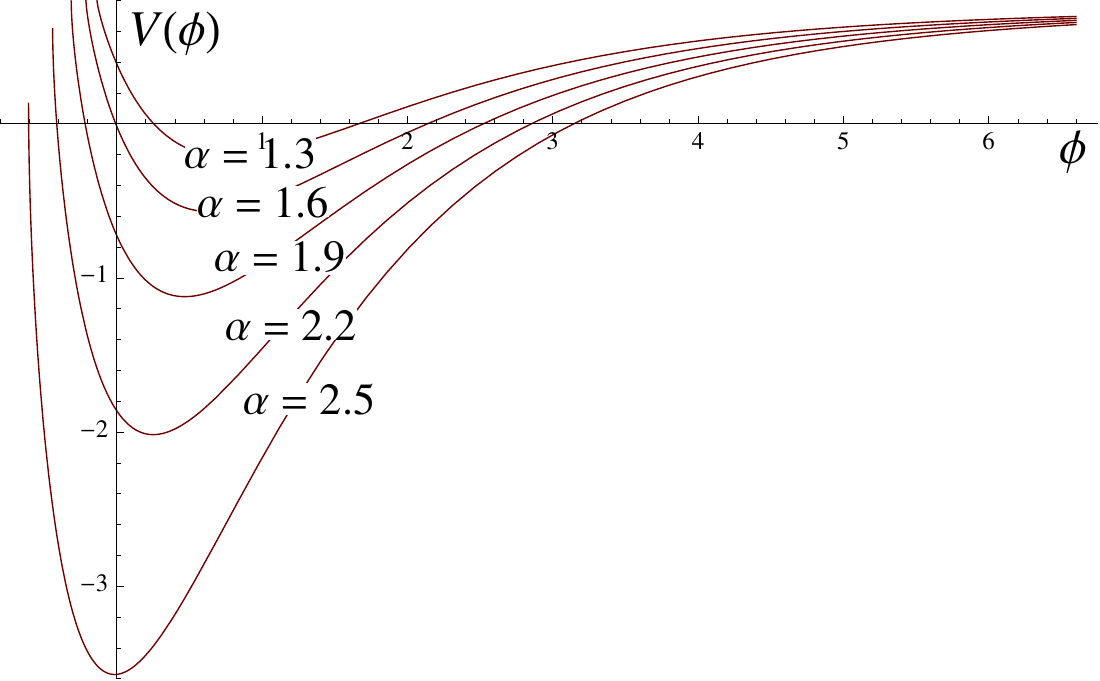}
\caption{Inflation potential for various values $\alpha$ and $\Lambda=1.4$.}
\label{f1}
\end{figure}

\section{inflation dynamics and primordial spectrum}
For $\Lambda\gg 1$ the potential behaves as  
$V_{\pm}(\phi) \sim \tfrac{1}{2}\,{e^{-2\sqrt{\frac{2}{3}}\kappa \phi}m^2\Lambda}/{\kappa^2}$ 
while for $\alpha\gg 1$ we have  $V_{\pm}(\phi) \sim  {\pm}\alpha^4 {e^{-2\sqrt{\frac{2}{3}}\kappa \phi}m^2}/{\kappa^2}$.
In both cases this implies $a_E (t)\sim t^{3/8}$ which does not provide inflation.

In order to study the inflation scenario in slow-roll approximation, we need to know the shape of the potentials $V_{\pm}(\phi)$. 
First we notice that, for all values $(\alpha,\Lambda)$, the potential 
has the plateau $V_{\pm}(\phi)=\tfrac{3 m^2}{4 \kappa^2}$ for $\phi\to\infty$. To verify slow-roll conditions inflation must 
start from $\phi > \mpl$ and then proceed from the right to left.
The behavior of the potential for $\phi\ll \mpl$, and thus the inflation scenario, strongly depends on the values $(\alpha,\Lambda)$. 
In particular $V_{\pm}(\phi)$ can either develop a minimum, or $V_{\pm}(\phi)\to -\infty$ for $\phi\to -\infty$. 
In this work we study the class of potentials such that the inflation ends after a finite number $N$ of e-folds, 
with a phase of parametric oscillations of the field $\phi$. 
In order to have a well defined reheating phase it is clear that the potential must have a minimum; 
furthermore it can be proved that, in our case, a {well defined exit from inflation} 
occurs only for potentials with a mininum $\phi_{\text{min}}$ such that $V(\phi_{\text{min}})\leq0$. 
These conditions are verified only for $V_+(\phi)$ if $\alpha\in[1,3]$ and $\Lambda\in[0,1.5]$ 
(the potential $V_-(\phi)$ can have a minimum for some special values of $(\alpha,\Lambda)$, 
but it is always $V(\phi_{\text{min}})>0$). 
In these cases the potential is depicted in \fig{f1} for various values of $\alpha$. 

In other words, although  for $\alpha$ and $\Lambda$ very close to zero \eq{effv}  is only a small modification 
of  the classical Starobinsky $R+R^2$ model, for $\alpha \in [1,3]$ and $\Lambda \in[0,1.5]$ 
the potential $V_+(\phi)$ develops a non-trivial minimum at {negative} 
values of the potential which makes our model significantly different from the original 
$R+R^2$ Starobinsky inflation.   In particular, as we shall see, after exit from inflation, 
the dynamics of the preheating phase
is characterized by a {\em lower limit} to $|\dot \phi|$ as it can be seen in \fig{f2} in the
$\phi$-$\dot \phi$ plane (see \cite{verno} for a general study of inflationary potentials with a {negative} value of 
the minimum).

Let us introduce the slow-roll parameters $\epsilon$ and $\eta$, 
\be 
\epsilon (\phi) \equiv \frac{1}{2\kappa^2} \left ( \frac{V'(\phi)}{V(\phi)} \right)^2, \quad \eta(\phi) = \frac{1}{\kappa^2} \left (\frac{V''(\phi)}{V(\phi)} \right) 
\ee
and define the number of e-folds before the end of inflation 
\be
\label{efol}
N=\int_{t_N}^{t_\text{end}} H(t) \,dt .
\ee
The amplitude of the primordial scalar power spectrum reads
\be
\label{ampli}
\Delta^2_{\cal R} = \frac{\kappa^4\, V(\phi)}{24 \pi^2\, \epsilon(\phi)}
\ee \\[0.1cm]
and, in this approximation, the spectral index $n_s$ and the tensor-to-scalar ratio $r$
are given by 
\begin{equation}
\begin{aligned}
& n_s= 1-6\, \epsilon (\phi_{i})+ 2\,\eta (\phi_{i}) \\[0.15cm]
& \r = 16\, \epsilon (\phi_{i})
\end{aligned}
\end{equation}

Where $\phi_{i}=\phi(t_N)$ is the value of the inflaton field $\phi(t)$ at the beginning of the inflation.
\begin{figure}[t]
\centering
\includegraphics[width=8cm]{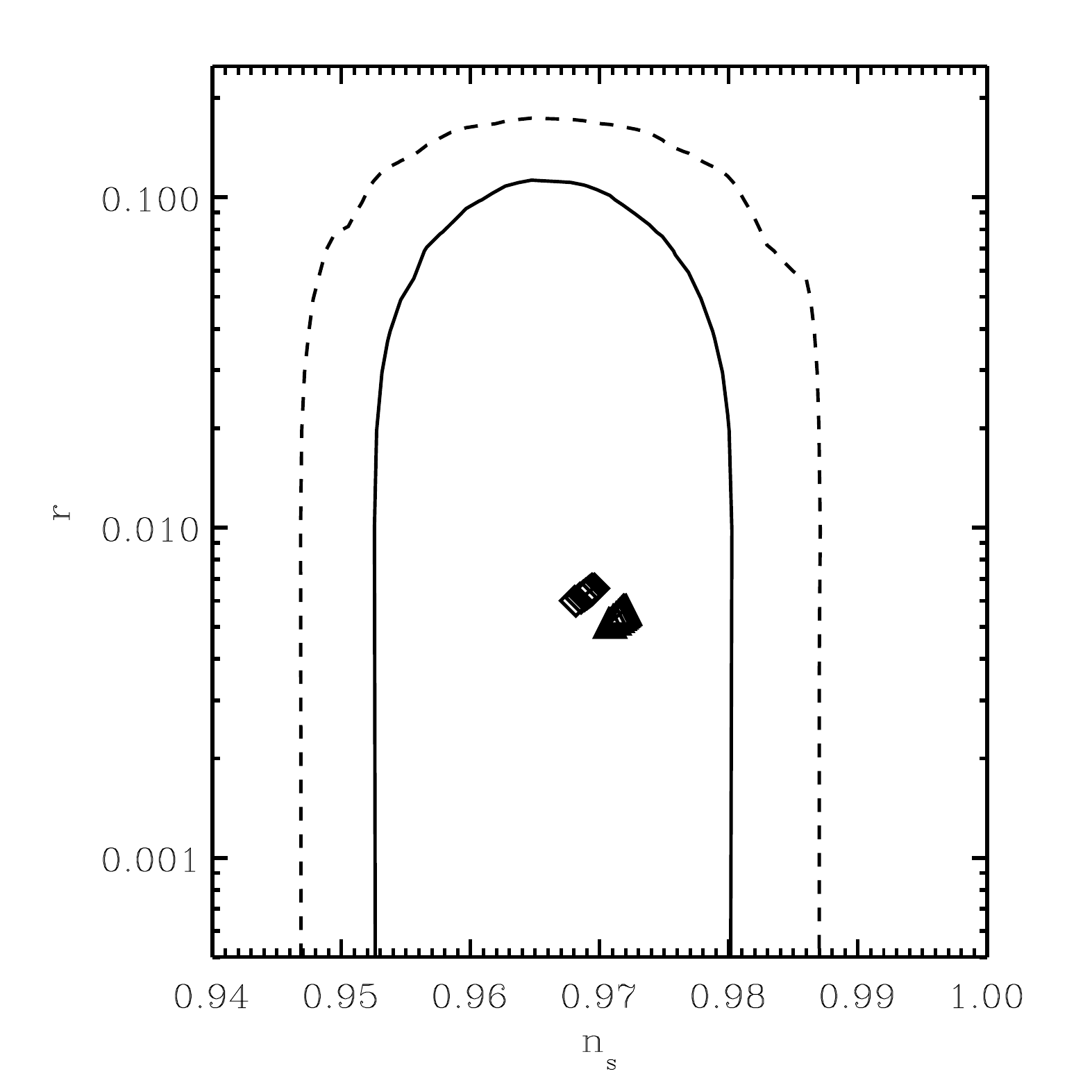}
\caption{\label{f3}
We compare the theoretical predictions in the $r$-$n_s$ plane for different values of $\alpha$ for the \planck collaboration
2015 data release for the TT correlation assuming $\Lambda$CDM + $r$ \cite{pla15}.
Triangles are for $N=55$ and squares for $N=60$ e-folds. Solid and dashed lines are the $1\sigma$ and $2\sigma$ confidence levels, respectively.
\label{pla}} 
\end{figure}

In our case it is not possible to analytically evaluate these expressions, and we must resort to a numerical estimation of the initial value
of the field which determines the final number of e-folds at the  end of inflation. 
As usual inflation is supposed to stop when the slow-roll condition is violated, $\epsilon(\phi)=1$. 
This condition determines the final  value $\phi_f=\phi(t_\text{end})$ of the inflaton field, 
while the initial value of the field $\phi(t)$ is obtained by inverting \Eq{efol}, once a number of e-folds is fixed. 
Our results are displayed in the following table:
\[
\begin{array}{@{\hspace{0.1cm}}c@{\hspace{0.3cm}}c@{\hspace{0.3cm}}c@{\hspace{0.3cm}}c@{\hspace{0.3cm}}c@{\hspace{0.3cm}}c@{\hspace{0.3cm}}c@{\hspace{0.3cm}}c@{\hspace{0.1cm}}}
\hline \\[-0.3cm]
\multicolumn{2}{c}{ \text{Cases}}  & \multicolumn{2}{c}{N=50} & \multicolumn{2}{c}{N=55}   & \multicolumn{2}{c}{ N=60} \\[0.1cm]
\hline \\[-0.3cm]
 \Lambda  & \alpha  & n_s & r & n_s & r & n_s & r \\[0.1cm]
\hline \\[-0.3cm]
 & 1.0 & 0.965 & 0.0069 & 0.968 & 0.0058 & 0.971 & 0.0050 \\[0.1cm]
0  & 1.8 & 0.966 & 0.0074 & 0.969 & 0.0063 & 0.972 & 0.0055 \\[0.1cm]
 & 2.6 & 0.967 & 0.0076 & 0.969 & 0.0065 & 0.972 & 0.0056 \\[0.1cm]
\hline \\[-0.3cm]
 & 1.0 & 0.965 & 0.0070 & 0.968 & 0.0059 & 0.971 & 0.0051 \\[0.1cm]
1  & 1.8 & 0.966 & 0.0074 & 0.969 & 0.0063 & 0.972 & 0.0055 \\[0.1cm]
 & 2.6 & 0.967 & 0.0076 & 0.969 & 0.0065 & 0.972 & 0.0056 \\[0.1cm]
\hline
\end{array}
\]
Confronting $n_s$ and $r$ with the '15 \planck data, our model agrees very well, as displayed in \fig{pla}.

According to \Eq{ampli} the normalization of the scalar power spectrum at the pivot scale  
$k_\ast=0.05\,\text{Mpc}^{-1}$ provides us with  $m\sim (1.5 \div 7) \cdot 10^{14}\,\text{GeV}$, 
depending on the value of $\alpha$.  

It should be stressed that, in our model, if $\Lambda<0$ there would be no exit from inflation with the standard reheating phase. 
As argued in \cite{dona} the presence of matter field could 
change the sign of $\Lambda$ depending on the number of Dirac, scalar and vector fields. 
We hope to address this issue in a future investigation. 

\section{Oscillatory phase after inflation}
After the end of inflation, the inflaton field $\phi$ begins to oscillate around the minimum $\phi_\text{min}$ 
of $V_+(\phi)$. To study this phase, we can do the following approximation
\begin{equation}
V_+(\phi) \sim V(\phi)=\frac{a}{2}\left[(\phi-\phi_\text{min})^2-b\right]
\end{equation}
Where $\phi_\text{min}(\alpha,\Lambda)$, $a(\alpha,\Lambda)={V}_+''(\phi_\text{min})$ and $b(\alpha,\Lambda)=-2\, V_+(\phi_\text{min})/{V}_+''(\phi_\text{min})$ depend on the values of $\alpha$ and $\Lambda$, and are given by the following expressions
\begin{equation*}
\phi_\text{min}(\alpha,\Lambda)=\frac{\sqrt{\frac{3}{2}}\left(3 \alpha ^3 \left(\alpha ^2-4\right)-32 \alpha  \Lambda +4 \left(\alpha ^2-6\right)|\alpha|^3\right)}{6 \alpha  \left(\alpha ^2-8\right) \left(\alpha ^2+2\right)-64 \alpha  \Lambda +8 \left(\alpha ^2-9\right)|\alpha|^3}
\end{equation*}
\begin{equation*}
a(\alpha,\Lambda)=\frac{48+18\, \alpha ^2-3 \,\alpha ^4+32 \,\Lambda -4\,\alpha ^3\, |\alpha|+36\,\alpha\,|\alpha|}{24} \;\,
\end{equation*}
\ba
b(\alpha,\Lambda)\, &&= \frac{8 \alpha  \left(15 \alpha ^4-3 \alpha ^6-96 \Lambda +8 \alpha ^2 (15+4 \Lambda )\right) |\alpha |}{\frac{8}{3} \left(48+18 \alpha ^2-3 \alpha ^4+32 \Lambda -4 \alpha  \left(\alpha ^2-9\right) |\alpha| \right)^2} \nonumber\\
&& + \: \frac{-25 \alpha ^8+132 \alpha ^6-384 \alpha ^2 \Lambda }{\frac{8}{3} \left(48+18 \alpha ^2-3 \alpha ^4+32 \Lambda -4 \alpha  \left(\alpha ^2-9\right) |\alpha| \right)^2} \nonumber\\
&& + \: \frac{48 \alpha ^4 (21+4 \Lambda ) -1024 \Lambda  (3+\Lambda )}{\frac{8}{3} \left(48+18 \alpha ^2-3 \alpha ^4+32 \Lambda -4 \alpha  \left(\alpha ^2-9\right) |\alpha| \right)^2} \nonumber
\ea
The time evolution of the field $\phi(t)$ is given by the Friedmann equation
\begin{equation}
\label{fried}
\ddot{\phi}(t)+3 \, H(t) \;\dot{\phi}(t)+V'(\phi(t))=0
\end{equation}
Where $H(t)$ is the Hubble constant
\ba
&& 3\,H(t)\, =3\left[\frac{1}{3}\left(\frac{1}{2}\dot{\phi}(t)^2+V(\phi (t))\right)\right]^{{1}/{2}} \nonumber\\
&& =\sqrt{\frac{3}{2}}\;\left[\dot{\phi}(t)^2+a\left(\phi (t)-\phi _{\min }\right){}^2-\text{ab}\right]^{{1}/{2}}
\ea
Putting $x(t)=\sqrt{a}\,(\phi(t)-\phi_\text{min})$ and $y(t)=\dot{\phi}(t)$, equation \eqref{fried} reads 
\begin{equation}
\label{apds}
\begin{cases}
\dot{y}=-\left[\frac{3}{2}\left(y^2+x^2-\text{ab}\right)\right]^{\frac{1}{2}}y-\sqrt{a}\: x \\ 
\dot{x}=\sqrt{a}\:y
\end{cases}
\end{equation}
The long time behavior of this dynamical system is mainly determined 
by the sign of $ab=-2\,V(\phi _{\min })$. If $ab\leq0$ (i.e. $V(\phi _{\min })\geq 0$) then the point $(0,0)$,
that is the minimum point $(\phi_\text{min},V(\phi_\text{min}))$, is an {attractive node}. 
If $ab>0$ (i.e. $V(\phi _{\min })< 0$) a {limit cycle} ($y^2+x^2=ab$) appears. 
Note that, in this sense, $ab=0$ is an {Hopf bifurcation point}.
\begin{figure}[t]
\centering
\includegraphics[width=7.7cm]{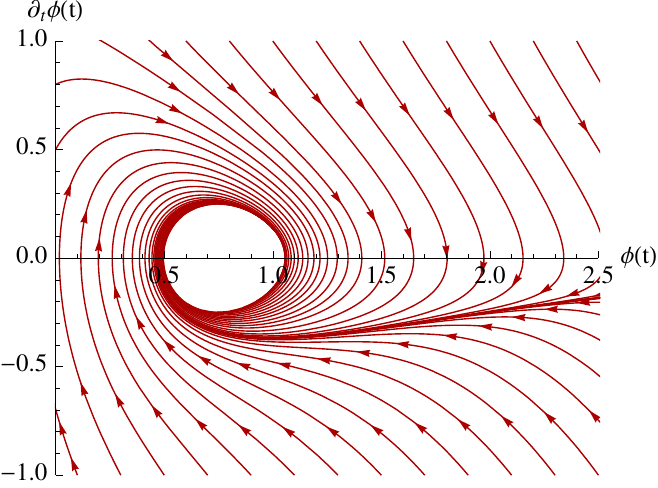}
\caption{\label{f2} 
Phase space evolution in the $\phi$-$\dot \phi$ space for $\alpha =1$ and $\Lambda=1$. Note the presence
of a limit cycle. }
\end{figure}
\begin{figure}[th]
\centering
\includegraphics[width=7.7cm]{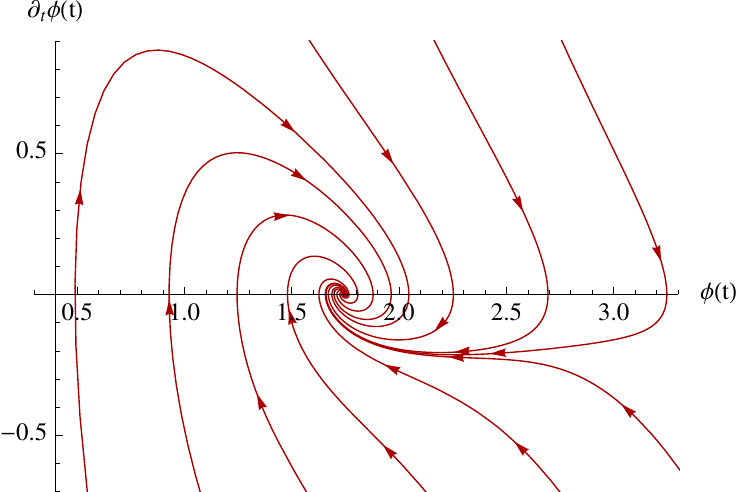}
\caption{\label{f4} 
Phase space evolution in the $\phi$-$\dot \phi$ space for $\alpha =1.5$ and $\Lambda=10$. This case arise when $V_+(\phi_\text{min})>0$. }
\end{figure}
A numerical study of the original Friedman equation \eq{fried} has confirmed our analytical findings. 
In particular we have obtained the phase diagrams relative to the cases $V_+(\phi_\text{min})<0$ (our case, \fig{f2}) and $V_+(\phi_\text{min})>0$ (\fig{f4}). 
The dynamical system analysis can be useful to determine the scale factor $a(t)$ after the inflation era, 
that is related to the Hubble constant by $H(t)=\frac{\dot{a}(t)}{a(t)}$. In particular, using the polar coordinates and then performing the method of averaging, we obtain 
\begin{equation}
a(t)=
\begin{cases}
\left[\sin\left(\sqrt{\frac{3}{8}\: |ab|} \;t\right)\right]^{2/3} & ab>0 \\
t^{2/3} & ab=0 \\
\left[\sinh\left( \sqrt{\frac{3}{8}\: |ab|} \; t\right)\right]^{2/3} & ab<0 \\
\end{cases}
\end{equation}
The scale factor $a(t)$, in the case $ab=0$ ($V(\phi_\text{min})=0$, Starobinsky model), 
describes the usual matter dominated era, while the solutions with $V(\phi_\text{min})\neq0$ are compatible 
with a matter dominated era only at the beginning of the oscillatory phase.
On the other hand a consistent treatment of the following reheating phase  must include the contribution of the
matter fields, an investigation which is beyond the scope of this work.\\[0.45cm]

\section{Conclusions}

In this work we have extended the idea of \cite{alfio12} by including in the renormalized flow the presence
of the additional relevant direction associated to the $R^2$ operator. We have approximated the flow of this operator
around the NGFP with its linear expression, an approximation which should capture the essential qualitative features
of the flow and allow an analytical treatment of the resulting non-linear $f(R)$ Lagrangian.\\[-0.1cm]

The most important point of our investigation is that our inflation model should significatly differ from the well
known Starobinsky model because it predicts a tensor-to-scalar ratio which is significantly higher, 
and an inflationary dynamics which is characterized by a limit-cycle behavior at the inflation exit. 
More important, our predictions are in agreement  with the latest \planck data which put important constraints on our model.\\[-0.1cm]

An  important limitation of this study is the simple tensorial structure of the effective Lagrangian which assumes  
a functional dependence of the $f(R)$ type. 
Quadratic operators like $R_{\mu\nu} R^{\mu\nu}$ are also associated to 
relevant directions around the NGFP and in principle their presence could dramatically change the inflation dynamics
and the generation of the primordial spectrum of fluctuations. 
On the other hand, our approach can easily be extended in order to include the contribution of these additional operators, 
whose critical exponents have been calculated in \cite{bema09}. 
We hope to address this issue in a future work.

\begin{acknowledgments}

The authors would like to thank Gian Paolo Vacca  and Sergey Vernov for important suggestions and comments.

\end{acknowledgments}

\bibliography{fr}

\end{document}